\documentclass[
aps,prl,twocolumn,showpacs,preprintnumbers,
amsmath,amssymb,groupedaddress,longbibliography,
]{revtex4-2}
\usepackage{hyperref}
\hypersetup{colorlinks=true, citecolor=blue, urlcolor=blue, linkcolor=blue,urlcolor=cyan}
\usepackage{color}
\usepackage{graphicx}
\usepackage{dcolumn}
\usepackage{braket}
\usepackage{amsthm}
\usepackage[caption=false]{subfig}

\def\be{\begin{equation}}
\def\ee{\end{equation}}
\def\ba{\begin{eqnarray}}
\def\ea{\end{eqnarray}}

 \def\ba{{\bar{\alpha}}}

\def\CM{{\mathcal{M}}}

\begin{document}

\title{Monitoring a de Sitter universe through an anti-de Sitter window}

\author{Sarah Hoback}
\affiliation{\it 
Jefferson Physical Laboratory, Harvard University, Cambridge, MA 02138, USA
}

\author{Daniel L. Jafferis}
\affiliation{\it 
Jefferson Physical Laboratory, Harvard University, Cambridge, MA 02138, USA
}

\author{Zixia Wei}
\affiliation{\it 
Jefferson Physical Laboratory, Harvard University, Cambridge, MA 02138, USA
}

\begin{abstract}

We propose that AdS$_3$ gravity coupled to dS$_2$ end-of-the-world branes is dual to a unitary holographic CFT$_2$ with non-unitary conformal boundary conditions. These boundary conditions have complex $g$-functions and appear in conjugate pairs. 
The associated gravitational path integral admits no real saddles, but does admit complex saddles.
We show that an AdS$_3$ black hole microstate with a dS$_2$ brane behind the horizon corresponds to the unitary time evolution of a pure CFT state prepared by a Euclidean path integral on a cylinder with such boundary conditions.
The construction predicts a boundary-condition-changing primary with $h=-c/8$, which resides in the boundary sector rather than the bulk spectrum and is therefore compatible with unitarity of the underlying CFT. 
This realizes dS holography as state preparation in a unitary AdS/CFT Hilbert space. 

\end{abstract}

\maketitle

{\bf 1. Introduction.---} 
How do you put an elephant in a refrigerator? This question has a logically simple answer: Open the door, put the elephant in, and close the door. However, the last step is technically nontrivial. 

In this paper, we answer a question which has a similar structure. How do you construct a de Sitter (dS) holography? Given the fact that holography in anti-de Sitter (AdS) spacetime is well-understood 
in terms of the conformal field theory (CFT) 
through the AdS/CFT correspondence \cite{Maldacena97}, a straightforward answer to this question is: Embed a dS spacetime into an AdS spacetime, and identify its counterpart in the CFT Hilbert space. Again, its first step is straightforward, but its last step is technically nontrivial.

Existing proposals for dS holography ranges from dS/CFT \cite{Strominger01,Witten01,Maldacena02} to various constructions where holographic screens are placed at timelike hypersurfaces inside dS \cite{AKST04, AHH11, GST18, Susskind21-2, NV23}. Here, we pursue a different approach by embedding dS in AdS and exploiting the power of AdS/CFT. Related attempts in this direction have appeared in \cite{FHMMR05,Mirbabayi20,AKTW20,BNU23,Wei24}.

Spacetimes that join AdS and dS regions have long been studied, and time-dependent classical solutions of this type exist in Einstein gravity coupled to a scalar field whose potential has both positive and negative minima \cite{FHMMR05}. The future $i^+$ boundary of the dS region is separated from the timelike AdS boundary by a horizon \cite{Penrose64,HR70} \footnote{An argument is that on a null congruence from the AdS timelike boundary there is initially contraction, and thus it cannot re-expand to reach $i^+$ of dS given the averaged null energy condition.}, so they can be described as an AdS and a dS black hole joined behind their horizons. 

Although closed dS universes (ie. the spatial slice is a closed manifold) have a large Hilbert space in perturbative quantum gravity \cite{Higuchi91,Higuchi91-2,MM08,CCGPR23}, there exists strong evidence that, non-perturbatively, the Hilbert space is one dimensional \cite{MV20,BKU20,BNU23,UWZ24,HUZ25,AAIL25}. 
This agrees with the proposal that the spacetimes with a dS region in the interior correspond to pure states of the CFT dual to AdS. An interpretation is that they correspond to a dS universe with some degrees of freedom (here a black hole) that are entangled with a reference (here an AdS black hole), which in the exact theory leads to a pure state of the reference system (here the CFT dual to AdS). An alternative direct argument is that by continuously deforming the data on an initial value slice, one can change whether a region of $i^+$ appears or if the dS region collapses. Since the latter type of spacetime seems to be clearly dual to a pure state at the AdS boundary, so must the former. 

\begin{figure}
    \centering
    \includegraphics[width=1.0\linewidth]{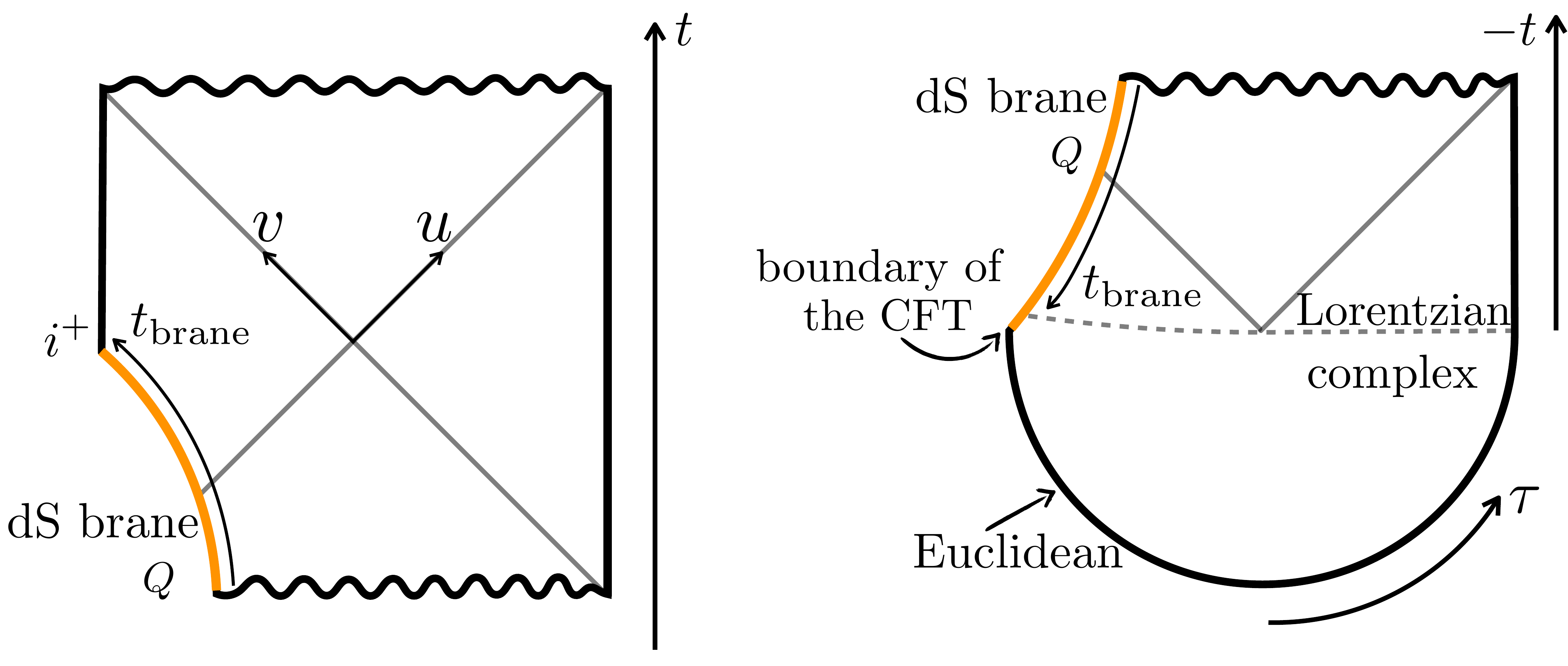}
    \caption{The left panel shows the Lorentzian geometry where a dS$_2$ EOW brane sits behind the event horizon of an AdS$_3$ BTZ black hole, where $t_{\rm brane}$ is the brane proper time and $t$ is the asymptotic boundary time. $u$ and $v$ are the two null directions appearing in the Kruskal coordinate. The right panel illustrates how such a Lorentzian geometry is obtained by analytic continuation from a complex geometry whose boundary is Euclidean. This will be shown to be dual to $e^{-itH_{\rm CFT}}e^{-\beta H_{\rm CFT}/2}|B\rangle$, where $\beta$ is the inverse temperature of the black hole and $|B\rangle$ is a non-unitary boundary state of the CFT in this Letter.}
    \label{fig:dS_behind_BTZ}
\end{figure}

We will investigate the related, but technically simpler scenario, of an end-of-the-world (EOW) brane whose intrinsic worldvolume describes an expanding dS universe. There are classical solutions in AdS gravity theories admitting the appropriate type of branes, and again the dS brane must be separated from the timelike AdS boundary by a horizon. 
See the left panel of Fig.\ref{fig:dS_behind_BTZ}. 
Note that the appearance of a component of the left AdS boundary is an artifact of exact rotational symmetry and the approximation of pure 3D gravity with a featureless EOW brane. The intersection of the EOW and the left AdS boundary is the start of a Cauchy horizon with infinite blue-shift for perturbations, and the expectation is that full generic solutions have the singularity pulled down to that location \cite{FHMMR05}.

These configurations can be interpreted as starting with a pair of large-$N$ matter CFTs' entangled state dual to the two-sided AdS black hole, and turning on dynamical de Sitter gravity on one boundary, in analogy with double holography \cite{AMMZ19} and braneworld setups \cite{RS99-1,RS99-2,KR00}.
In the full non-perturbative theory, the result becomes a pure state of the remaining non-gravitational CFT, that semi-classically was a purifying reference system for the large-$N$ matter theory on the dS boundary.

In the construction we will present, all states ultimately live in the Hilbert space of conventional CFTs, with standard unitary time evolution. The physics of dS spacetime is encoded in the procedure to produce those states. We will show that the associated dual objects (here a non-unitary conformal boundary condition) live off the real section in a complexification of spacetime. Therefore our construction is agnostic to the ultimate fate of the $i^+$ region of the real Lorentzian dS that is tied to its (meta)-stability. Once we use the dual dS ingredients to produce a pure state of the CFT living on the AdS boundary, it will have a real Lorentzian gravity dual whose evolution is governed by whatever are the laws of quantum gravity in that universe. 

~\par

{\bf 2. dS$_2$ brane behind AdS$_3$ black hole horizon.---}
Before moving to our holographic construction, let us present an AdS black hole geometry with a dS EOW brane behind the horizon. We start with a two-sided BTZ black hole \cite{BTZ92} in AdS$_3$, whose metric can be written as 
\begin{align}\label{eq:BTZ}
    ds^2 = -\frac{4}{(1+uv)^2}du\,dv + r_h^2 \frac{(1-uv)^2}{(1+uv)^2} d\phi^2\,,
\end{align}
in the Kruskal coordinates. We have set the AdS radius to 1, and $r_h$ is related to the black hole temperature $\beta^{-1}$ as $r_h = 2\pi \beta^{-1}$. The asymptotic boundary is $uv = -1$, and the black hole singularity is $uv =1$ in this coordinate.

The hypersurface 
\begin{align}\label{eq:brane_profile_Lorentzian}
    u(v) = -\frac{1+\alpha v }{\alpha + v}\,, u<v\,,
\end{align}
where $0<\alpha <1$ is a constant,
describes a dS$_2$ universe with dS radius $R_{\rm dS_2} = {\sqrt{1-\alpha^2}}/\alpha$ with induced metric 
\begin{align}
    ds^2|_Q = - \frac{4(1-\alpha^2)}{\alpha^2(1-v^2)^2} dv^2 + r_h^2 
    \left[\frac{\alpha\left(1+v^2\right)+2v}{\alpha (1-v^2)}\right]^2
    d\phi^2 \,.
\end{align}
If we consider the $u>-(1+\alpha v )/(\alpha+v)$ part as the physical region, we obtain an AdS$_3$ black hole with a dS$_2$ EOW brane inside. See the left panel of Fig.\ref{fig:dS_behind_BTZ} for a sketch of the Penrose diagram. 

Note that now we have just written down a geometry without speaking out what theory has such a solution. In the following, we write down such a theory and identify its CFT dual.

~\par

{\bf 3. Setup, obstruction, and resolution.---} We consider AdS$_3$ Einstein gravity coupled to 2D EOW branes with tension $T$, described by the Euclidean action 
\begin{align}\label{eq:action}
    I_{E} =& -\frac{1}{16\pi G_N} \int_{\CM} \sqrt{g}\left(R+2\right) \nonumber - \frac{1}{8\pi G_N} \int_Q \sqrt{h}(K-T) \nonumber \\
    &- \frac{1}{8\pi G_N} \int_\Sigma \sqrt{\gamma} \,B + I_{\rm c.t.}\,. 
\end{align}
The first term is the Einstein-Hilbert term, where $(\CM,g_{\mu\nu})$ denotes the 3D manifold, $R$ is the Ricci scalar, and $G_N$ is the Newton constant. Note that we have set the cosmological constant to $\Lambda = -1$ such that the AdS radius is 1. The second term is the Gibbons-Hawking term on the EOW brane $Q$, where $K_{ab}$ is the extrinsic curvature and $h_{ab}$ is the induced metric. $T$ is the tension term, for which we set $T>1$. The third term is the Gibbons-Hawking term on the asymptotic boundary $\Sigma$, where $B_{ij}$ is the extrinsic curvature and $\gamma_{ij}$ is the metric. $I_{\rm c.t.}$ is the counter term.
We define the extrinsic curvature such that the normal vector points outwards.

We fix the geometry at the asymptotic boundary $(\Sigma,\gamma_{ij})$, but not that in the bulk $\CM$ or on the brane $Q$. The equation of motion (EOM) obtained from varying the bulk geometry is the Einstein equation $R_{\mu\nu} - \frac{1}{2} R g_{\mu\nu} - g_{\mu\nu} = 0$, and the EOM on the brane $Q$ reads 
\begin{align}\label{eq:NBC}
    K_{ab} - (K-T)h_{ab} = 0\,.
\end{align}
This is often called the Neumann boundary condition (NBC) in literature. 

It is straightforward to see that the geometry shown in Eq.\eqref{eq:BTZ}\eqref{eq:brane_profile_Lorentzian} and the left panel of Fig.\ref{fig:dS_behind_BTZ} is a solution of this theory with $T$ and $\alpha$ related as  
\begin{align}\label{eq:alpha_T}
    \alpha = \frac{\sqrt{T^2-1}}{T}\,.
\end{align}
In general, the tension $T>1$ guarantees that the worldvolume of $Q$ is locally dS$_2$ on shell. 

Let us then consider the CFT$_2$ dual to the AdS$_3$ theory given by \eqref{eq:action} via the Gubser-Klebanov-Polyakov-Witten (GKP-W) dictionary \cite{GKP98,Witten98}.

For a given Euclidean manifold $\Sigma$, we consider the gravitational path integral (GPI)
\begin{align}\label{eq:GPI}
    Z_{\rm AdS}[\Sigma] = \int_Q \mathcal{D}h_{ab} \int_{\partial\CM = Q\cup \Sigma} \mathcal{D}g_{\mu\nu} \exp\left(-I_E\right)\,,
\end{align}
where we sum over all the 3D configurations $\CM$ (including the brane configurations on $Q$) compatible with $\Sigma$, and interpret it as a CFT partition function on $\Sigma$ following the GKP-W dictionary 
\begin{align}
    Z_{\rm CFT}[\Sigma] = Z_{\rm AdS}[\Sigma]\,.
\end{align}
Since the full GPI is not in general computable, in practice, we perform saddle point approximations. 

To see why the CFT$_2$ dual of a dS$_2$ EOW brane is tricky, let us look at a simple on-shell geometry. Consider the hyperbolic ball (Euclidean AdS$_3$) parameterized by $r = [0,\infty)$, $\theta \in \left[-\frac{\pi}{2}, \frac{\pi}{2}\right]$, and $\phi \in [0,2\pi)$ with metric 
\begin{align}\label{eq:hyperbolic_ball}
    ds^2 = dr^2 + \sinh^2 r \left( d\theta^2 + \cos^2 \theta \,d\phi^2\right)\,. 
\end{align}
Taking $r < r_* \equiv {\rm arccoth}(T)$ as the the physical region by identifying $r = r_*$ as $Q$ gives an on-shell configuration, but it does not have an asymptotic boundary, i.e. $\Sigma = \varnothing$. 

In fact, there does not exist any Euclidean solution in which the dS$_d$ brane intersects with the AdS$_{d+1}$ asymptotic boundary \cite{AKTW20,Wei24}, and hence it is highly nontrivial how the dS$_d$ branes should be sourced in CFT, and what choice of $\Sigma$ can make the GPI dominated by a configuration with a dS$_d$ brane. 
This situation is very different from an AdS$_d$ brane: it intersects the AdS$_{d+1}$ asymptotic boundary, which makes it straightforward to relate it to CFT$_d$ with boundaries or defects as in \cite{Takayanagi11,FTT11}. Previous work has tackled this problem by considering $\Sigma$ to be a complexified manifold \cite{AKTW20}, a non-orientable manifold \cite{Wei24}, or a Lorentzian manifold with spacelike boundaries \cite{AKRTW21,AKRTW22,FKKT25}, but neither of them admits a unitary-time-evolution interpretation on the CFT side.

A key observation to be made in this Letter is that there exist complex solutions where the dS brane $Q$ intersects the asymptotic boundary $\Sigma$. When we fix $\Sigma$ to be a Euclidean manifold with boundaries, e.g. a disk, and perform the Euclidean GPI, these complex solutions serves as saddle points of the GPI, though it does not admit any Euclidean saddle points. 

In the following, we fix $\Sigma$ to be a 2D Euclidean manifold with boundaries and look for complex solutions compatible with $\Sigma$. We use them to compute the corresponding CFT quantities holographically, and explain what kind of boundary CFT (BCFT) they are dual to. 

In the end, we will see that the dS-brane-behind-BTZ geometry is dual to a BCFT defined on a cylinder $\Sigma$, but let us start from the most simple case where $\Sigma$ is a disk. 

~\par
{\bf 4. Disk $\Sigma$ \& non-unitary boundary condition.---} Let $\Sigma$ be a disk. A simple solution compatible with it can be constructed as follows. Consider the Euclidean AdS$_3$ ball Eq.\eqref{eq:hyperbolic_ball} and perform the coordinate transformation, 
\begin{align}
    \cosh\,r = \cosh\,\eta\, \cosh{\,\rho}\,,~~~~\sinh\,r\,\sin\,\theta = \sinh\,\rho \,.
\end{align}
The metric then reads 
\begin{align}
    ds^2 = d\rho^2 + \cosh^2\rho \left(d\eta^2+\sinh^2\eta \,d\phi^2\right)\,.
\end{align}
In this coordinate, each $\rho = {\rm const.}$ gives an AdS$_2$ slice. The asymptotic boundary of AdS$_3$, which is a sphere, consists of two hemispheres $\rho \rightarrow \pm \infty$, respectively. On the other hand, $\eta \rightarrow \infty$ is the asymptotic boundary for each AdS$_2$ slice.  

\begin{figure}
    \centering
    \includegraphics[width=1.0\linewidth]{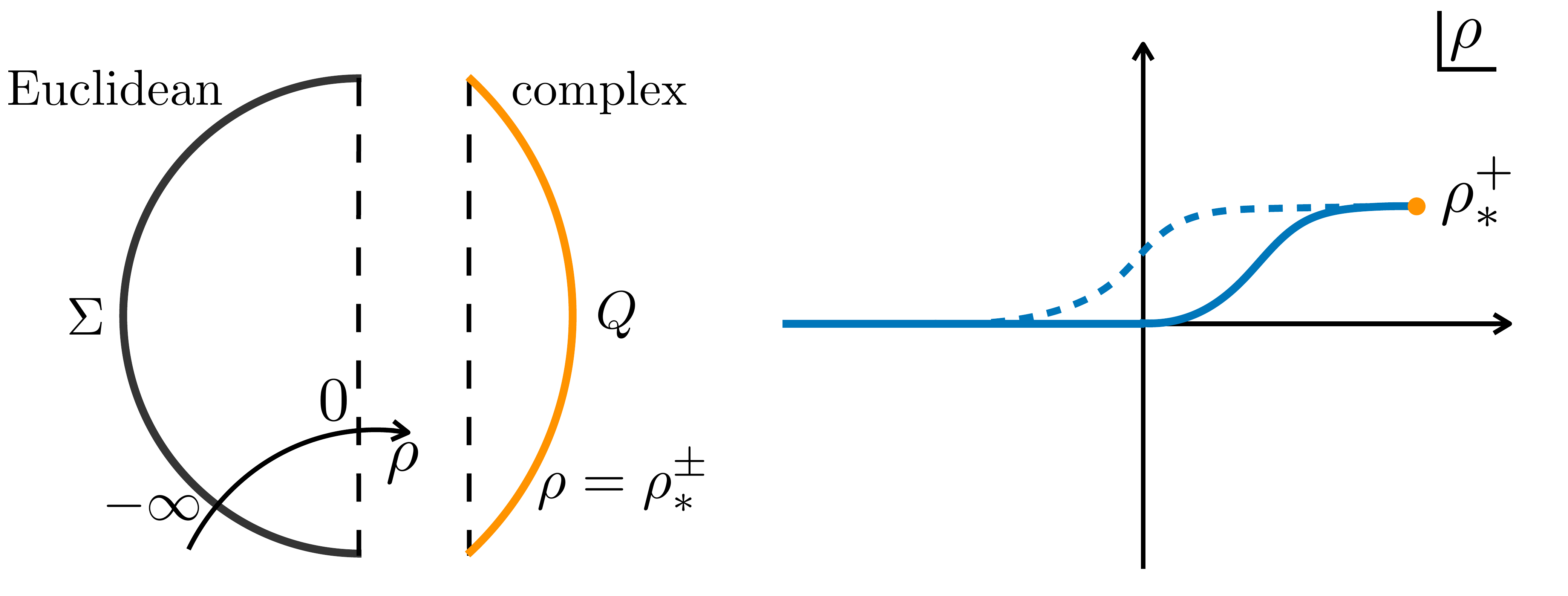}
    \caption{The left panel shows a 2D slice of the half AdS$_3$ ball configuration $\rho\in(-\infty,0]$ and the complex disk wedge interpolating between $\rho = 0$ and $\rho_*^\pm$. A solution for $T>1$ is obtained by pasting them, which can be represented as a contour (solid blue) on the complex $\rho$-plane (right panel). 
    Other contours (e.g. dashed blue) smoothly deformed from it can be considered as a complex analytic continuation, and can be used equivalently.}
\label{fig:disk_solution}
\end{figure}

Although we are considering $T>1$, in which case there does not exist Euclidean solutions, it is instructive to start by considering the Euclidean half ball configuration $\rho \in (-\infty, 0]$ as shown in Fig.\ref{fig:disk_solution}.
This would be a solution if $T$ was zero, with $\rho=0$ identified as $Q$ and $\rho \rightarrow -\infty$ as $\Sigma$. More generally, the Euclidean configuration given by $\rho \in (-\infty, \rho_*]$ would be a solution for $T = \tanh \rho_*$ when $0<T<1$ \cite{AKTW20}. We can divide this into $\rho \in (-\infty, 0]$ and  $\rho \in [0, \rho_*]$, where the former part is a solution for $T=0$, i.e. a half ball, and the latter part is called a disk wedge (since its topology is $D^2\times I$). In general, if one wants to construct a solution for $0<T<1$ and evaluate its action, it turns out to be useful to first construct a solution for $T=0$ and then glue a wedge region to it. This is called the wedge trick \cite{WWW25}. 

Now let us generalize this construction to $T>1$. Complexify the variable $\rho$ \footnote{Complex metrics on real coordinates can equivalently be represented as real metrics on complexified coordinates. We adopt the latter representation here for convenience.}, and define a region running from $\rho = 0 $ to 
\begin{align}\label{eq:rho_star}
     \rho ={\rm arccoth}\, T \pm i\frac{\pi}{2}\, \equiv \rho^{\pm}_* ,
\end{align}
to be a complex disk wedge, as shown in Fig. \ref{fig:disk_solution}. It is straightforward to see that attaching such a complex disk wedge to the half ball $\rho \in (-\infty, 0]$ gives a complex solution for $T>1$, where $\rho = \rho_*^\pm$ is identified as the EOW brane $Q$. Indeed, the intrinsic metric on $Q$ turns out to be $-$AdS$_2$, i.e. AdS$_2$ with $(-,-)$ signature, which solves Eq.\eqref{eq:NBC}, the EOM of dS$_2$ brane. 

Let us then evaluate its action. Divide it into the $\rho \in (-\infty, 0]$ part and the wedge part. For the $\rho \in (-\infty, 0]$ part, let us put the bulk IR cutoff at $r = r_{\infty}$ such that the cutoff shape is a hemisphere, which is related to a disk by a Weyl transformation. Substitute $B = 2 \coth r_{\infty}$, $R=-6$ to the action integral and subtract the divergent terms by a counter term, 
\begin{align}
    I_\text{half ball} = &-\frac{1}{16\pi G_N} \times 2\pi \int_0^{r_\infty} dr \sinh^2 r\times (-6+2) \nonumber\\
    &-\frac{1}{8\pi G_N} \times 2\pi \sinh^2 r_\infty \times 2\coth r_{\infty} +I_{\rm c.t.} \nonumber\\
    = & -\frac{1}{8G_N} \,.
\end{align}
For the disk wedge, with consistent cutoff $\eta = r_\infty$, 
\begin{align}\label{eq:disk_wedge}
    I^\pm_\text{disk wedge} = &\frac{1}{4\pi G_N} \int_0^{\rho_*^{\pm}} d\rho \cosh^2 \rho \times 2\pi\int_{0}^{r_{\infty}}d\eta \sinh \eta \nonumber \\ 
    &- \frac{T}{8\pi G_N} \cosh^2 \rho_*^{\pm} \times 2\pi\int_{0}^{r_{\infty}}d\eta \sinh \eta \nonumber \\
    &- \frac{1}{8\pi G_N} \cosh r_{\infty} \times 2\pi \int_{0}^{\rho_*^\pm} d\rho \nonumber \\
    =& -\frac{\rho_*^{\pm}}{4G_N}\,,
\end{align}
where the three terms are the bulk term, the brane term, and the AdS boundary term, respectively. 
Accordingly, on the corresponding CFT side, we have  
\begin{align}
    &\log\frac{Z_{\rm CFT}[S^2/\mathbb{Z}_2]}{\sqrt{Z_{\rm CFT}[S^2]}} = -I_\text{half ball} - I^\pm_\text{disk wedge} + I_\text{half ball} \nonumber \\
    =& \frac{\rho_*^\pm}{4G_N} = \frac{c}{6} \rho_*^{\pm} \equiv \log g_{\pm}, 
\end{align}
where $c$ is the central charge of the corresponding CFT given by the Brown-Henneaux relation $1/4G_N = c/6$ \cite{BH86}. $g_{\pm}$ is interpreted as the $g$-function associated to the solution where the EOW brane sits at $\rho = \rho^{\pm}_*$, respectively. The two $g$-functions are related by 
\begin{align}
    g_- = g_+^*\,.
\end{align}
Note that we treat the two solutions as corresponding to two different boundary conditions in BCFT, rather than two saddle points contributing to the same BCFT. In fact, in the two solutions, the EOW brane approaches the asymptotic boundary at different angles, which should be considered as a part of the asymptotic data. 

Let us use $B$ and $\bar{B}$ to denote the boundary condition with $g$-function $g_+$ and $g_-$, respectively. More precisely, on the CFT side, a general conformal boundary condition $B$ can be represented as a boundary state $|B\rangle \in \mathcal{H}_{\rm CFT}$ satisfying $(L_n-\bar{L}_{-n})|B\rangle$ = 0, 
where $L$'s are the Virasoro generators. A general solution of it can be written as 
\begin{align}\label{eq:boundary_state}
    | B \rangle = \sum_j d_j |j\rangle\rangle\,,
\end{align}
where $|j\rangle\rangle$ is the Ishibashi state \cite{Ishibashi88} associated to the bulk primary labeled by $j$. The coefficient for the identity Ishibashi state $d_\mathbb{I}$ is identical to the $g$-function. The same boundary condition, when viewed as a bra state, reads $\langle B | = \sum_j  d_j \langle \langle j|$. For such a boundary condition $B$, we define its conjugate boundary condition $\bar{B}$ as the one whose associated boundary state is 
\begin{align}
    | \bar{B} \rangle = \sum_j d_j^* |j\rangle\rangle\,.
\end{align}
Accordingly, $\langle \bar{B} | = \sum_j  d_j^* \langle \langle j|$. Therefore, we have 
\begin{align}
    |B\rangle^\dagger = \langle \bar{B} |\,,
\end{align}
where $\dagger$ is the usual Hermitian conjugate associated with $\mathcal{H}_{\rm CFT}$ defined from the Virasoro representation theory. A special class of conformal boundary conditions whose dual open-string theory is unitary is often considered \cite{Cardy89}. Such a unitary boundary condition $B_{\rm uni}$ has real $d_j$'s and satisfies $|B_{\rm uni}\rangle^{\rm \dagger} = \langle B_{\rm uni}|$. In this sense, the complex AdS solutions discussed above are dual to disk BCFT with non-unitary boundary conditions. By contrast, conformal boundary conditions appearing in usual AdS$_3$/BCFT$_2$ with AdS$_2$ branes \cite{Takayanagi11,FTT11} are unitary.

~\par

{\bf 5. Cylinder $\Sigma$ with $BB$ boundaries.---} 
Let us then proceed to consider the case where $\Sigma$ is a cylinder, which has two boundaries. Let the interval direction of $\Sigma$ be parameterized by $\tau \in [-L/2,L/2]$ and the $S^1$ direction be parameterized by $\phi \in [-\pi,\pi)$ with $\phi \sim \phi + 2\pi$. We first consider imposing $B$ on both boundaries. We refer to this case as $BB$-cylinder.

There are two solutions in this case. To describe them, let us again apply the wedge trick \cite{WWW25}. The first solution has two disconnected pieces of the EOW brane $Q$, which can be constructed as follows. First, the configuration with the metric 
\begin{align}
    ds^2 = d\chi^2 + \cosh^2 \chi \,d\tau^2 + \sinh^2 \chi\,d\phi^2\,,
\end{align}
interpolating between $\tau=-L/2$ to $\tau = L/2$ would be a solution if $T$ was zero, with $\tau = \pm L/2$ identified as two disconnected pieces of the EOW brane $Q$. This is exactly a half of the thermal AdS geometry when regarding $\tau$ as the Euclidean time direction. Attaching a complex disk wedge described in Eq.\eqref{eq:rho_star} and Fig.\ref{fig:disk_solution} to $\tau = -L/2$ and another one to $\tau =L/2$, we obtain a complex solution compatible with $BB$-cylinder $\Sigma$ with two disconnected pieces of the brane $Q$. Let us call this the disconnected solution. Its on-shell action reads 
\begin{align}
    I^{BB}_{\rm dis} &= I_\text{half thermal AdS} + 2 I^+_\text{disk wedge} \nonumber \\
    &= -\frac{L}{8G_N} - \frac{\rho_*^+}{2G_N} = -\frac{cL}{12} - \frac{c\rho_*^+}{3}\,, 
\end{align}
which is complex.

The other solution has only one connected piece of the EOW brane $Q$, constructed as follows. First, consider the Euclidean BTZ geometry  
\begin{align}\label{eq:BB_BTZ}
    ds^2 = d\chi^2 + \left(\frac{\pi}{L}\right)^2\sinh^2 \chi \,d\tau^2 + \left(\frac{\pi}{L}\right)^2\cosh^2 \chi\,d\phi^2\,,
\end{align} 
with $\tau \in (-L,L]$, $\tau \sim \tau + 2L$. 
The inverse temperature is $\beta = 2L$. 
With the coordinate transformation 
\begin{align}
    \sinh \rho = \sinh \chi\, \sin \frac{\pi\tau}{L},\, \tanh \eta = \tanh \chi \cos \frac{\pi\tau}{L},\, \varphi = \frac{\pi}{L}\phi
\end{align}
we have 
\begin{align}
    ds^2 = d\rho^2 + \cosh^2 \rho \left(d\eta^2 + \cosh^2 \eta \,d\varphi^2\right)\,,
\end{align}
where each $\rho = {\rm const.}$ is locally AdS$_2$ with the cylinder topology. Its asymptotic boundary, which is a torus, consists of two cylinders $\rho \rightarrow \pm \infty$. $\eta \rightarrow \pm\infty$ are the two  asymptotic boundaries for each AdS$_2$ slice. The configuration interpolating $\rho \in (-\infty, 0]$ would be a solution if $T$ was zero, with $\rho=0$ identified as the EOW brane $Q$. Now, define a region running from $\rho = 0 $ to $\rho = \rho_*^+$ to be a complex cylinder wedge. Its action can be evaluated in a similar way to \eqref{eq:disk_wedge}, which turns out simple
\begin{align}
    I_\text{cylinder wedge}^+ 
    =0\,.
\end{align}
Attaching a complex cylinder wedge to the half BTZ $\rho \in (-\infty, 0]$ gives a complex solution for $T>1$, where $\rho=\rho_*^+$ is the EOW brane $Q$. Let us call it the connected solution. Its on-shell action reads
\begin{align}
    I^{BB}_{\rm con}  = I_\text{half BTZ} + I_\text{cylinder wedge}
    =
    -\frac{\pi^2}{8G_NL} = -\frac{\pi^2c}{12L}.
\end{align}
The dominating saddle point is then given by the one whose action's real part is minimal. 

The holographic partition function for the $BB$-cylinder is in general complex. This is because the corresponding BCFT computes $\braket{B|e^{-LH_{\rm CFT}}|B}$, which 
\begin{align}
    \left(\braket{B|e^{-LH_{\rm CFT}}|B}\right)^{\dagger} = \braket{\bar{B}|e^{-LH_{\rm CFT}}|\bar{B}} \neq \braket{B|e^{-LH_{\rm CFT}}|B}\,.
\end{align}
The dual open-string Hamiltonian $H_{\rm open}^{BB}$ satisfies 
\begin{align}
    {\rm Tr}\,e^{-2\pi H_{\rm open}^{BB}} = \braket{B|e^{-LH_{\rm CFT}}|B}\,,
\end{align}
and is non-Hermitian. This is again because we do not require unitarity for the open-string theory \cite{Cardy89}. 

At the $L\ll 1 $ limit, the connected solution dominates, and the holographic energy stress tensor turns out to be 
\begin{align}
    T_{\phi\phi}^{\rm (hol)} = -T_{\tau\tau}^{\rm (hol)} =-\frac{1}{16\pi G_N} \left(\frac{\pi}{L}\right)^2 = -\frac{c}{24}\frac{\pi}{L^2}\,,
\end{align}
which is the same as the energy density for the CFT ground state defined on an $S^1$ with circumference $2L$. This is consistent with the fact that $\bra{B}$ and $\ket{B}$ are described by the same boundary condition and can be connected by the identity operator on the boundary. 

The $\bar{B}\bar{B}$-cylinder case is completely parallel, so we will not repeat its analysis here. The $B\bar{B}$ case is fundamentally different, which is our next focus.

~\par 
{\bf 6. Cylinder $\Sigma$ with $B\bar{B}$ boundaries.---} 
Let us then consider the $B\bar{B}$-cylinder, where $B$ is imposed on one boundary and $\bar{B}$ is imposed on the other. 

Again, we find two complex solutions. One is disconnected and is constructed analogously to the $BB$-cylinder case. The difference is that, instead of attaching two identical complex disk wedges, we attach one wedge associated with $g_+$ and the other with $g_-$. The action is
\begin{align}
    I^{B\bar{B}}_{\rm dis} &= I_\text{half thermal AdS} +  I^+_\text{disk wedge} +  I^-_\text{disk wedge} \nonumber \\
    &= -\frac{L}{8G_N} - \frac{{\rm arccoth}\,T}{2G_N} = -\frac{cL}{12} - \frac{c }{3}{\rm arccoth}\,T\,, 
\end{align}
which is real. 

The other solution, which is a connected configuration with only one connected piece of $Q$, turns out to be very different. Start with the Euclidean BTZ geometry  
\begin{align}\label{eq:BBbar_BTZ}
    ds^2 = d\chi^2 + \left(\frac{2\pi}{L}\right)^2\sinh^2 \chi \,d\tau^2 + \left(\frac{2\pi}{L}\right)^2\cosh^2 \chi\,d\phi^2\,,
\end{align} 
with the inverse temperature $\beta = L$, which is twice different from Eq.\eqref{eq:BB_BTZ} considered in the $BB$-case. Complexify $(\chi,\tau)$ and perform the coordinate transformation 
\begin{align}
    u = e^{i\frac{2\pi}{L}\tau} \tanh \frac{\chi}{2}\,,~~~v = -e^{-i\frac{2\pi}{L}\tau} \tanh \frac{\chi}{2}\,,
\end{align}
the metric turns 
\begin{align}\label{eq:complex_Kruskal}
    ds^2 = -\frac{4}{(1+uv)^2}~du~dv + \left(\frac{2\pi}{L}\right)^2 \frac{(1-uv)^2}{(1+uv)^2} ~d\phi^2,
\end{align}
This is nothing but a complexified Kruskal metric \eqref{eq:BTZ} with $\beta = L$. 

\begin{figure}
    \centering
    \includegraphics[width=1.0\linewidth]{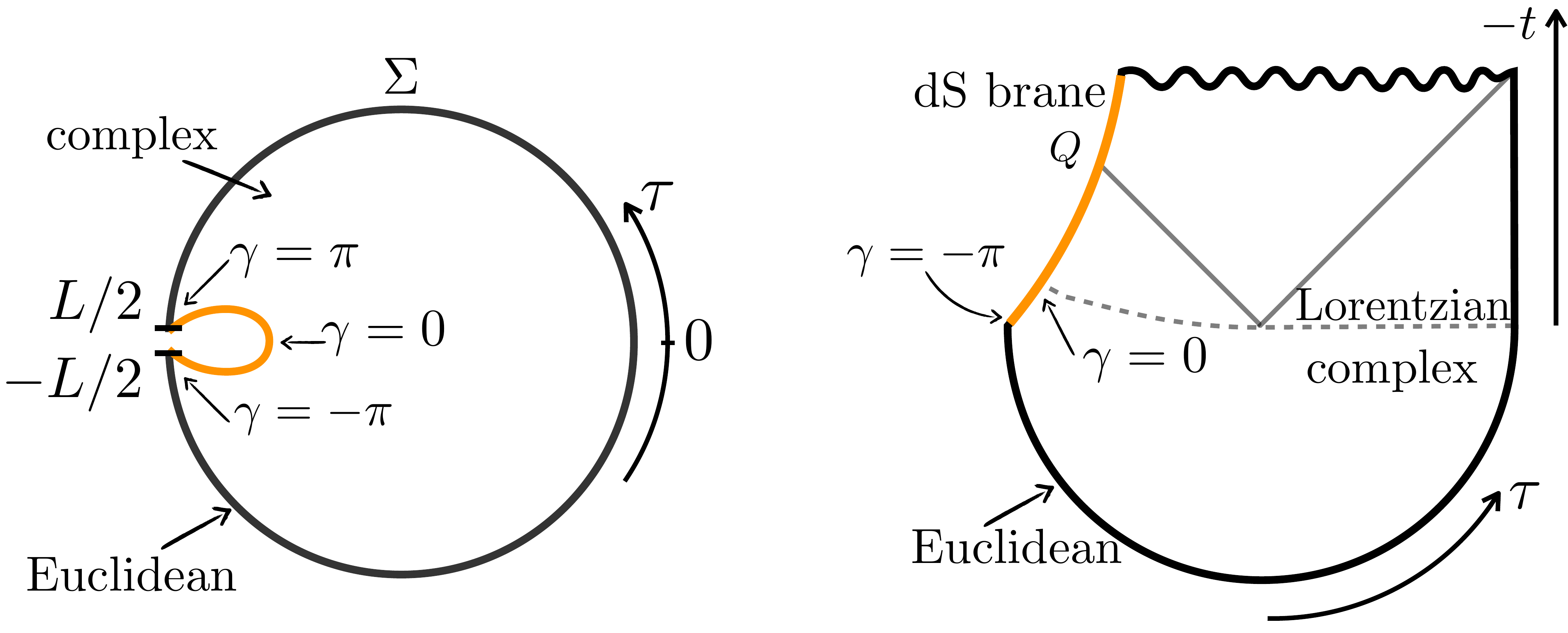}
    \caption{The left panel illustrates the $\phi={\rm const.}$ slice of the connected solution for $B\bar{B}$-cylinder $\Sigma$. The right panel shows its analytic continuation to Lorentzian signature. }
\label{fig:BBbar_solution}
\end{figure}

Let us introduce a real parameter $\gamma\in(-\pi,\pi)$ and a small constant $\varepsilon$ which acts as a cutoff \footnote{In fact, as we will see, it is not necessary but convenient to take the constant $\varepsilon$ to be small.}, then 
\begin{align}\label{eq:complex_brane_profile}
    (u,v) = \left(- \frac{1+\alpha-\alpha\varepsilon (1+e^{-i\gamma})}{1+\alpha-\varepsilon (1+e^{-i\gamma})}, 1-\varepsilon(1+e^{-i\gamma})\right)\,.
\end{align}
where $\alpha$ is Eq.\eqref{eq:alpha_T}, 
describes a 2D (complex) locus which solves complexified brane trajectory equation Eq.\eqref{eq:brane_profile_Lorentzian}. On the other hand, the Euclidean AdS boundary is given by
\begin{align}\label{eq:Euclidean_AdS_boundary}
    (u,v) = \left(e^{i\frac{2\pi}{L}\tau}, -e^{-i\frac{2\pi}{L}\tau}\right),~~~~\tau \in [-L/2,L/2]\,.
\end{align}
Therefore, both the complex brane profile Eq.\eqref{eq:complex_brane_profile}
and the Euclidean AdS boundary Eq.\eqref{eq:Euclidean_AdS_boundary} approaches the point $(u,v) = (-1,1)$ twice from different directions. Let us identify $\gamma \rightarrow \pm \pi$ as $\tau \rightarrow \pm L/2$.
Then the brane $Q$ and the asymptotic boundary $\Sigma$ intersects twice at different angles, as sketched in the left panel of Fig.\ref{fig:BBbar_solution}. Any smooth interpolation between the two serves as a complex bulk solution for the $B\bar{B}$-cylinder. 

A crucial point is that $\varepsilon$ can be taken to be arbitrarily small. As a result, we can construct such a bulk solution by subtracting an arbitrarily small complex bubble from the BTZ configuration, and its action reads 
\begin{align}
   I^{B\bar{B}}_{\rm con} = -\frac{\pi^2}{2G_NL}= -\frac{\pi^2c}{3L}, 
\end{align}
which is identical to that of a Euclidean BTZ geometry with inverse temperature $\beta = L$, and real.

The dominating saddle point is again given by the one with the minimal action. The holographic partition function is real, which is consistent with 
\begin{align}
    \left(\braket{\bar{B}|e^{-LH_{\rm CFT}}|B}\right)^{\dagger} = \braket{{\bar{B}}|e^{-LH_{\rm CFT}}|B} \,.
\end{align}
In the $L\ll 1 $ limit, the connected solution dominates. The holographic energy stress tensor turns out to be 
\begin{align}
    T_{\phi\phi}^{\rm (hol)} = -T_{\tau\tau}^{\rm (hol)} =-\frac{1}{16\pi G_N} \left(\frac{2\pi}{L}\right)^2 = -\frac{c}{6}\frac{\pi}{L^2}\,. 
\end{align}
This indicates that there exists a nontrivial boundary condition changing (bcc) operator connecting $B$ and $\bar{B}$. Letting $h_{\rm bcc}$ be its (holomorphic) conformal dimension, we have 
\begin{align}
    \left(h_{\rm bcc}^{B\bar{B}} - \frac{c}{24}\right) \frac{\pi}{L^2} = -\frac{c}{6}\frac{\pi}{L^2}~\Longleftrightarrow~ h_{\rm bcc}^{B\bar{B}}= -\frac{c}{8}\,.
\end{align}
The negativity of $h_{\rm bcc}^{B\bar{B}}$ is a signature that the dual open-string theory with $B\bar{B}$ boundaries is non-unitary. Note that the bcc operator does not live in the spectrum of the original CFT, which remains unitary. However, it is essential for the holographic BCFT to have such a bcc operator to accommodate a dS EOW brane in the bulk. 

~\par 
{\bf 7. Analytic continuation and global quench.---} Let us consider the analytic continuation of the $B\bar{B}$ cylinder setup discussed above. Thanks to the time reflection symmetry with respect to $\tau =0$, it realizes a density matrix $\rho = e^{-LH_{\rm CFT}/2}|B\rangle \langle \bar{B}| e^{-L H_{\rm CFT}/2}$ at $\tau = 0$. 
At high temperature, i.e. small $L$, the analytic continuation $\tau = it$ of the connected solution is dual to the unitary time evolution of the pure state 
\begin{align}\label{eq:quenched_pure_state}
    \ket{\Psi(t)} = e^{-itH_{\rm CFT}}e^{-LH_{\rm CFT}/2}\ket{B}\,,
\end{align}
living in $\mathcal{H}_{\rm CFT}$. Continuing the whole bulk to Lorentzian signature, we obtain the dS-behind-BTZ configuration as shown in Fig.\ref{fig:dS_behind_BTZ} and the right panel of Fig.\ref{fig:BBbar_solution}.

From the CFT perspective, our setup generalizes the Calabrese--Cardy global quench \cite{CC05}. There, the initial state is a unitary boundary state satisfying $\ket{B}^{\dagger}=\bra{B}$, and the Euclidean cutoff $L/2$ sets the effective temperature to $\beta^{-1}=(2L)^{-1}$. Our analysis suggests a broader class of quenches with $\ket{B}^{\dagger}=\bra{\bar B}\neq\bra B$, for which the bcc operator connecting $B$ and $\bar B$ enhances the temperature. This is realized in our $B\bar B$-cylinder analysis. The effective temperature is $\beta^{-1}=L^{-1}$, twice higher than that in \cite{CC05}, and is controlled by a bcc operator of dimension $h_{\rm bcc}^{B\bar B}=-c/8$. Correspondingly, while the conventional global quench is naturally dual to AdS EOW branes \cite{HM13}, our construction is dual to a dS EOW brane behind the BTZ horizon. Related generalized quenches are realized using an AdS$_2$ brane coupled to an imaginary scalar field \cite{MNT26}. See also \cite{TWW26} for related setups in free-fermion/boson CFT and \cite{KTWY23} for temperature enhancement in generic product states.

~\par
{\bf 8. Nonunitary boundary conditions in unitary CFT.---} In unitary CFTs, not all the linear combinations of Ishibashi states in the form of \eqref{eq:boundary_state} are considered to be physical. In fact, unitarity of the open-string theory is often required to construct physical boundary conditions \cite{Cardy89}. This leaves the question of whether we are allowed to consider non-unitary boundary condition as discussed above. Indeed, non-unitary boundary conditions are easy to construct in simple CFTs. For example, in the free compact boson CFT, the Dirichlet boundary condition fixes the field configuration at the boundary to a real value. One can consider analytically continuing these boundary fields to complex values to construct non-unitary boundary conditions as in, e.g. \cite{GRW01,AJ22}. Another example is non-unitary boundary conditions in unitary Liouville CFT, obtained from analytically continued FZZT boundary conditions, as explicitly noted in \cite{Teschner00}. Physical nonunitary boundary conditions can be also explicitly constructed in lattice systems \cite{TWW26}. 

One might wonder if there is a tension between the open sector Cardy formula and complex $g$-function, since an asymptotic spectral density $\rho(h) \sim g^2 \left(\frac{c}{48 h^3}\right)^{\frac{1}{4}} \exp\left(\pi \sqrt{\frac{c h}{3}}\right)$ would not be real for real $h$. However, the derivation is from matching $\int dh \rho(h) e^{-\pi \beta h}$ with $Z(\beta) \sim g^2 e^{\pi c /12\beta}$ at high temperatures, so any contour in the $h$ plane that results in the same integral suffices. The requirement that physical operators even in a non-unitary system have real spectral density, $\rho(h) dh$, is satisfied if asymptotically $\textrm{Im}\sqrt{h} \sim - \sqrt{\frac{12}{\pi c}} \arg(g)$.

~\par
{\bf 9. Conclusion and Discussion.---} 
How do you put a dS EOW brane in AdS/CFT? One includes such branes in the gravitational action and performs the GPI with an asymptotic boundary $\Sigma$ that is itself a manifold with boundary. The dS EOW branes then appear in complex saddle points. The dual theory is a BCFT built from a unitary parent CFT, but endowed with non-unitary boundary conditions whose $g$-functions are complex.

We illustrated this construction for disk and cylinder $\Sigma$. In this framework, a black hole microstate with a dS EOW brane behind the horizon is dual to unitary time evolution of a pure CFT state, prepared by a Euclidean path integral with boundary condition $B$. We emphasize that the non-unitarity of the boundary condition does not affect the unitarity of the parent CFT, in which the Hilbert space and states are discussed. 

Our construction provides a realization of dS holography through an AdS/CFT window. The dS region is encoded in the machinery that produces an ordinary CFT state that evolves unitarily in time. Our construction conceptually sits between dS/CFT \cite{Strominger01,Witten01,Maldacena02}, where dS gravity is dual to a non-unitary CFT, 
and formalisms with timelike holographic screens inside dS space \cite{AKST04, AHH11, GST18, Susskind21-2, NV23}, in the following sense. In our construction, the dS$_2$ gravity on the EOW brane may be viewed as associated with non-unitary CFT$_1$ data. However, rather than treating this as an autonomous quantum theory, we regard it as a conformal boundary condition that prepares a state in the Hilbert space of AdS$_3$/CFT$_2$. The holographic screen is timelike, but it is the standard AdS boundary, lying outside of the dS universe rather than inside it. 

More broadly, our construction gives an answer to the question of what to do with dS/CFT if you had one: 
instead of interpreting it as a quantum theory itself, we can consider it as a parasite that makes a state living in someone else's Hilbert space. Refer to \cite{GJ26} for a related JT-gravity setup in which an ensemble dual is found for a dS region connected to an AdS region by a domain wall.

Several directions remain open. First, since the construction gives a CFT bootstrap type characterization of dS branes, 
one can invert the logic to discover a dS brane given a concrete boundary condition with appropriate properties in a known holographic CFT. Second, the Lorentzian evolution of the dS brane, including possible decays or bubble nucleations, can in principle be probed by complicated bulk reconstruction. 
Physics localized on the brane could be probed by light boundary operators dual to brane-localized fields \cite{KSSTW23}. For example, one may formulate and compute braneworld scattering observables directly from BCFT correlators. 
Finally, our construction identifies a new class of black hole microstates whose detailed properties remain to be investigated. 

In the end, we have succeeded in ``putting the elephant into the refrigerator'' and ``closing the door.'' 
This may remain difficult for ordinary elephants and refrigerators, but dS branes are more accommodating: they can enter through complex saddles.

~\par 
\textbf{Acknowledgments.---}
We are grateful to Ping Gao, Simeon Hellerman, Juan Maldacena, Yu Nakayama, Eva Silverstein, Douglas Stanford, Tadashi Takayanagi, Seiji Terashima for helpful discussions and comments.
ZW thanks the organizers and participants of ``Holography and Quantum Information" at YITP and ``de Sitter Holography Meets Non-Hermitian Quantum Matter" at RIKEN, where this work was presented.  
SH is supported by the NSF GRFP Fellowship. The work of DLJ is supported in part by Simons Investigator in Physics Award MP-SIP-0001737 and U.S. Department of Energy
grant DE-SC0007870.
ZW is supported by the Society of Fellows at Harvard University. 

\bibliography{dS_AdS}

\end{document}